\documentclass[times,doublespace]{simauth}

\usepackage{moreverb,multirow}

\usepackage[dvips,colorlinks,bookmarksopen,bookmarksnumbered,citecolor=red,urlcolor=red]{hyperref}

\newcommand\BibTeX{{\rmfamily B\kern-.05em \textsc{i\kern-.025em b}\kern-.08em
T\kern-.1667em\lower.7ex\hbox{E}\kern-.125emX}}

\def\volumeyear{2010}

\def\bSig\mathbf{\Sigma}

\DeclareMathOperator{\logit}{logit}

\begin{document}

\runninghead{N. C. Brownstein et al.}

\title{Parameter estimation in Cox models with missing failure
  indicators and the OPPERA study}

\author{Naomi C. Brownstein\affil{a,b,c}, Jianwen Cai\affil{c}, Gary D. Slade\affil{c}, and Eric Bair\affil{c,d}\corrauth\footnotemark[2]}

\address{\affilnum{a}Ion Cyclotron Resonance Program, National High
  Magnetic Field Laboratory, Florida State University, Tallahassee FL,
  U.S.A. \\
  \affilnum{b}Department of Statistics, Florida State University,
  Tallahassee FL, U.S.A \\
  \affilnum{c}Department of Biostatistics, University of North
  Carolina at Chapel Hill, Chapel Hill NC, U.S.A.\\
  \affilnum{d}School of Dentistry, University of North Carolina at
  Chapel Hill, Chapel Hill NC, U.S.A.}

\corraddr{Eric Bair, CB \#7450, School of Dentistry, Chapel Hill, NC
  27599-7450, U.S.A.}

\cgs{The OPPERA study was supported by NIH/NIDCR grant
  U01DE017018. Naomi Brownstein was supported by NIH/NIEHS grant
  T32ES007018 and NSF Graduate Research Fellowship Program grant
  0646083. Jianwen Cai was supported by NIH/NCI grant P01CA142538 and
  NIH/NIEHS grant R01ES021900. Eric Bair was supported by NIH/NIDCR
  grant R03DE023592, NIH/NIEHS grant P30ES010126, and NIH/NCATS grant
  UL1TR001111.}

\begin{abstract}
In a prospective cohort study, examining all participants for
incidence of the condition of interest may be prohibitively
expensive. For example, the ``gold standard'' for diagnosing
temporomandibular disorder (TMD) is a physical examination by a
trained clinician. In  large studies, examining all participants in
this manner is infeasible. Instead, it is common to use questionnaires
to screen for incidence of TMD and perform the ``gold standard''
examination only on participants who screen positively. 
Unfortunately, some participants may leave the study before receiving
the ``gold standard'' examination. Within the framework of survival
analysis, this results in missing failure indicators.  Motivated by
the Orofacial Pain: Prospective Evaluation and Risk Assessment
(OPPERA) study, a large cohort study of TMD, we propose a method for
parameter estimation in survival models with missing failure
indicators. We estimate the probability of being an incident case for
those lacking a ``gold standard'' examination using logistic
regression. These estimated probabilities are used to generate
multiple imputations of case status for each missing
examination that are combined with observed data in appropriate
regression models. The variance introduced by the procedure is
estimated using multiple imputation. The method can be used to
estimate both regression coefficients in Cox proportional hazard
models as well as incidence rates using Poisson regression. We
simulate data with missing failure
indicators and show that our method performs as well as or better than
competing methods. Finally, we apply the proposed method to data from
the OPPERA study.
\end{abstract}


\maketitle

\footnotetext[2]{E-mail: ebair@email.unc.edu}

\section{Introduction}
\label{s:intro}
Time-to-event analyses are frequently conducted in medicine, actuarial
science, and numerous other fields of applied science. There is a
well-developed set of survival analysis methods implemented in
standard software. 
Semi-parametric methods, such as the Cox
proportional hazards model, 
allow robust estimation of the effects of covariates on the hazard function.
However, these methods
require the analyst to know 
the failure status of each participant, which may not always be available.


In some cases the outcome of interest may be difficult to ascertain. For
example, in oncology studies, researchers may want to differentiate
between deaths due to cancer and deaths due to car accidents or other
unrelated causes. Investigators may easily record the mortality of all
subjects, but it may be extremely difficult or costly to find out
exactly why each subject died. One possible solution to this problem
is  delayed event adjudication \cite{kosorok}. This means that possible
cases are not identified immediately but screened
using simple methods that may have poor
sensitivity or specificity. Later, the
screened candidate cases are re-examined using a more precise, but
also more costly and
time-consuming, method to determine the true event status.


The study that motivates our work is Orofacial Pain: Prospective Evaluation
and Risk Assessment (OPPERA), a prospective cohort study to
identify risk factors for the onset of temporomandibular disorders (TMD).
Each (initially TMD-free) OPPERA study participant was followed for a
median of 2.8 years
to identify cases of first-onset TMD. However, it was
impractical to perform a physical examination on every participant. 
It would also have been inefficient given that most study participants
did not develop the condition. Instead, this ``gold standard'' examination was performed
only on participants with positive screens on 
a quarterly screening questionnaire 
that was designed to assess recent orofacial pain \cite{bairOpperaII}.
However, some participants with positive screens 
were lost to follow-up before receiving the ``gold standard'' examination.
Thus a time-to-event analysis would have some participants
with missing failure indicators.

Previous research indicates that when a subset of the failure
indicators are missing, one can obtain more accurate estimates of the
parameters of interest by using appropriate tools to estimate these
missing values \cite{kosorok,magaret,dodd}. Cook and Kosorok
\cite{kosorok} 
estimate parameters in Cox proportional hazard models with missing
failure indicators by weighting observations 
according to their probability of being a true case.
They show that the estimators are
consistent and asymptotically normally distributed.
However, the standard error of their proposed estimate cannot be easily
obtained using existing software without bootstrapping.
For the OPPERA data, a separate Cox model was calculated
for each putative risk factor of interest,
including approximately three thousand genetic markers.
Consequently, applying this method to the OPPERA genetic data
would be computationally intractable.


%

In the OPPERA study, the likelihood that a participant with a positive
screen
was examined was weakly associated with 
demographic variables such as gender, race, or socioeconomic status
\cite{bairOpperaII}.
This indicated that the failure indicators
in the OPPERA study were not missing completely at random (MCAR).
Application of
models that  assume MCAR failure indicators may result in biased
estimates of hazard ratios for covariates of interest.
More importantly, a participant's
responses to their screening questions are predictive of whether or not they are an
incident case of TMD. 
This setting presents statistical challenges, which require care in
order to avoid bias and maintain efficiency.
Additionally, incidence rate estimates are desired, and
none of the methods currently available allow for estimation of the incidence rate.
There is a clear need for new methodology to effectively answer
the research questions of the OPPERA study.


In this paper, we propose a method
for parameter and variance estimation in Cox regression models with
missing failure indicators.  The motivating data set is introduced in section
\ref{s:opperaintro}.  We describe our method in section
\ref{s:model}.  In section \ref{s:sims}, we report the results of
simulations.  Finally, in section \ref{s:inf} we apply our
method to the OPPERA study. We conclude
with a discussion in section \ref{s:discuss}.

\section{Motivating Data Set: The OPPERA Study}\label{s:opperaintro}
OPPERA is a prospective cohort study designed to identify risk factors
for first-onset TMD.
A total of 3,263 initially TMD-free subjects were recruited at four study sites
between 2006 and 2008.
TMD status was confirmed by physical examination of the jaw joints
and muscles using the Research Diagnostic Criteria for TMD
\cite{DworkinRDC}, which is the gold standard for diagnosing TMD.

Upon enrollment in the study, each OPPERA participant was evaluated for
a wide variety of possible risk factors for TMD, including
psychological distress, previous history of painful conditions, and
sensitivity to experimental pain. For a brief overview of the risk factors
of interest in the OPPERA study, see Section \ref{opperaappendix} in
the Supporting Information. 
See Ohrbach et al. \cite{OhrbachClinical},
Fillingim et al. \cite{FillingimPsych}, Greenspan et
al. \cite{GreenspanQST}, Maixner et al. \cite{MaixnerAutonomic}, and
Smith et al. \cite{SmithGenetics} for a complete description of the
baseline measures that were collected in OPPERA.

After enrollment, each
participant was asked to complete questionnaires to evaluate recent
orofacial pain once every three months.
These questionnaires
(hereafter referred to as ``screeners'') evaluated
the frequency and severity of pain in the orofacial region during
the previous three months. 
The purpose of the screener was to identify participants who
were likely to have recently developed TMD.
For a complete description of the screener, see
Slade et al. \cite{SladeQHU}. Participants with a positive screen
were asked to undergo
a follow-up physical examination by a clinical expert 
to diagnose presence or absence of TMD.



Of the 3,263 subjects, 2,737 filled out at least 1 screener, and the
remaining 521 did not fill out any screeners.  The total number of screeners was 
26,666. 
There were 717 positive screeners, 486 (about 68\%) of which were followed
by a clinical examination. As reported in Bair et al. \cite{bairOpperaII}, case classifications made by one examiner (hereafter, ``Examiner \#4'') were deemed unreliable
because the examiner diagnosed a much higher percentage of individuals with TMD compared to other examiners.
We therefore set all of Examiner \#4's physical examination findings to be missing and
imputed them using the methods in this paper.  This left 404 positive screeners (56\%)
resulting in valid clinical exams.


\section{Model}\label{s:model}
\subsection{Notation and Assumptions}
Assume there are $n$ independent participants. For each participant
$i$ ($i=1,\ldots,n$),
let $C_i$ and $T_i$ denote the potential times until censoring and failure,
respectively, let $V_i=\min(T_i,C_i)$, $\Delta_i=I(T_i\leq C_i)$.
Let $Z_i$ a $p\times1$ vector of covariates measured at baseline 
and let $X_i$ 
be a $q\times1$ vector
of covariates measured at the time of the putative event. We assume the hazard 
for participant $i$ follows a Cox proportional hazards model
\begin{equation}\label{coxprophaz}
\lambda(t|z_i)=\lambda_0(t)\exp(\beta^\prime z_i)
\end{equation}
where $\lambda_0(t)$ is an unspecified baseline hazard function.
%
Let $\xi_i$ denote the
indicator that $\Delta_i$ is observed.
We observe $(V_i,\xi_i)$ for $i=1,\ldots,n$ and $\Delta_i$ when
$\xi_i=1$.

In the OPPERA study, $V_i$ is the length of time for participant $i$ between
enrollment in the study and either of two events
\begin{enumerate}
\item a 
screener which resulted in a diagnosis of incident TMD
\item the last-completed screener before loss-to-follow-up.
\end{enumerate}
Note that participants with a positive screen do not fill out
additional screeners until they are examined, so $V_i$ will be the
time until the positive screen for a participant who has a positive
screen but is never examined. If participant $i$ had a positive screen
and subsequently was diagnosed with TMD, 
then $\Delta_i=1$. If participant $i$ either had a negative screen on
the last quarterly screener
before loss-to-follow up or a positive screen and was diagnosed to be
free of TMD, then $\Delta_i=0$.
If participant $i$ had a positive screen on the last screener 
but was not examined, then $\Delta_i$ is missing
and $\xi_i=0$. 
The putative risk factors
for TMD that were assessed at enrollment 
are denoted by the
vector $Z_i$. Responses to the screener 
for participant $i$ at time $V_i$ are denoted by the vector $X_i$. For OPPERA, we
also define $Q_i=1$ if participant $i$ has a positive screen on their
final screener and $Q_i=0$ otherwise.

We assume the failure indicators are missing at random (MAR) as follows:
\begin{equation} \label{marequation}
  P(\xi_i=0|X_i,Z_i,V_i,\Delta_i,Q_i=1) = P(\xi_i=0|X_i,Z_i,V_i,Q_i=1)
\end{equation}
In other words, the probability of having a missing failure indicator
may depend on measured factors, but it
does not depend on whether or not an event occurred. We will describe
how to estimate the probability in (\ref{marequation}) in Section
\ref{estevent} and then show how to use this estimate to impute the
missing event indicators in Section \ref{mi}.

\subsection{Estimating Event Probabilities} \label{estevent}
We model 
the probability that participant $i$ with a missing
failure indicator is a case 
by a logistic
regression model based on $X_i$ and $V_i$:
%
\begin{equation}\label{logisticmodel}
P(\Delta_i=1|X_i,V_i,Z_i,\xi_i=0,Q_i=1)=\frac{\exp(\alpha^\prime
  X_i+\gamma^\prime Z_i+\eta V_i)}{1+\exp(\alpha^\prime X_i+\gamma^\prime Z_i+\eta V_i)}I(Q_i=1)
\end{equation}

That is, we 
estimate the probability of examiner-diagnosed TMD in a participant
who was not examined as intended. (Here $I(x)$ denotes an indicator
function.) The probability was estimated using 
the time between enrollment and their last positive screener 
as well as their answers on that screener.
Then, for those individuals who had a positive screen on the last
screener
(i.e. those with $Q_i=1$) and were not examined, the
estimated probability of being a case is estimated by (\ref{logisticmodel})
with the parameters replaced by their respective estimates based on individuals
who were examined.

Note that this also assumes that there is one observation per subject,
which may not be the case in practice. For example, if some
participants had a positive screen on more than one screener and are
examined at least once, then we have multiple observations per
participant. In that case, fitting a generalized linear mixed effects
logistic regression model rather than a standard logistic regression
model could account for correlations between the responses of the same
participant. However, only a small number of participants in the
OPPERA study were examined multiple times after positive screeners, so
we simply discarded all but the most recent screener when analyzing
the OPPERA data (thereby avoiding this problem of repeated
observations).

%
\subsection{Multiple Imputation}\label{mi}
One popular method for handling missing data 
is multiple imputation.  
For a comprehensive review on multiple imputation, see Rubin \cite{rubinmi}.
Our imputation procedure is as follows:
\begin{enumerate}
\item Estimate the coefficients $\alpha$, $\gamma$, and $\eta$ in
  (\ref{logisticmodel}). We used a Bayesian model where $\alpha$,
  $\gamma$, and $\eta$ had a prior distribution that was Cauchy with
  center 0 and scale 2.5.
\item For each observation with a missing failure indicator, sample
  from the posterior distribution of $\alpha$, $\gamma$, and $\eta$
  to obtain an estimate of the probability that an event occurred for
  each such observation. \label{predictedprobs}
\item Generate a Bernoulli random variable with success probability equal
  to the predicted probability found in step (\ref{predictedprobs}). \label{imputemiss}
\item Combine the raw data and imputed data from step (\ref{imputemiss}) to
  form a completed data set.
\item Fit the Cox proportional hazards model to the completed data set.
\item Record each parameter estimate $\hat{\beta}_j$ and covariance
  matrix $\hat{U}_j$. \label{record}
\item Repeat steps (\ref{imputemiss})-(\ref{record}) for a total of $m$
  times, where $m$ is the desired number of imputations.
\end{enumerate}
Next, we combine all of the estimates.  The average
parameter estimate is
\begin{equation}\label{mimean}
\bar{\beta}=\frac{1}{m}\sum_{j=1}^m\hat{\beta}_j,
\end{equation}
the within-imputation variance estimate is
\begin{equation}\label{miparm}
\bar{U}=\frac{1}{m}\sum_{j=1}^m\hat{U}_j,
\end{equation}
and the between-imputation variance \begin{equation}
\hat{B}=\frac{1}{m-1}\sum_{j=1}^m (\hat{\beta}_j-\bar{\beta})(\hat{\beta}_j-\bar{\beta})^\prime.
\end{equation}
Finally, the estimated covariance matrix is
\begin{equation}\label{mivar}
\hat{\mbox{Var}}(\bar{\beta})=\bar{U} + \left(1+\frac{1}{m}\right)\hat{B}.
\end{equation}
It can be shown that $\bar{\beta}/\hat{\mbox{Var}}(\bar{\beta})$ is
approximately $t$ distributed with degrees of freedom
\begin{equation} \label{midf}
  (m-1) \left( 1+ \frac{m \bar{U}}{(m+1) \hat{B}} \right)
\end{equation}
(\ref{mivar}) and (\ref{midf}) can be used to compute confidence
intervals for the multiply imputed parameter estimate $\bar{\beta}$.
%


\subsection{Estimation of Incidence}
Previous sections of this paper described how to estimate hazard ratios
in the presence of missing failure indicators.
It may also be of interest to estimate incidence rates for the same
event using Poisson regression instead of Cox regression.
For example, one of the aims of the OPPERA study
is to estimate the incidence rate of first-onset TMD.

In order to estimate incidence rates, we estimate the case probabilities
as described previously based on participants who had a positive
screen and were examined.
Then we impute case status as described in section \ref{mi} for those
who had a positive screen but were not examined.
However, in this case we
fit Poisson regression models, rather than Cox models, to the completed data sets.
Finally, we calculate the incidence rate based on the estimates of the
regression coefficients in the Poisson model.
Specifically, we use the data from imputation $j$ to fit the model
\begin{equation}
\log({E(\Delta_{ij}|X_i,Z_i,V_i)})=\mu+\tau^\prime X_i+\lambda^\prime
Z_i+\log({V_i})
\end{equation}
where $\Delta_{ij}$ denotes the $j^{th}$ imputation for observation $i$,
$j=1,\ldots,m$. We combine the $m$ imputations using equation (\ref{mimean})
and
\begin{equation}
\bar{\mu}=\frac{1}{m}\sum_{j=1}^m\hat{\mu}_j.
\end{equation}
The estimated incidence rate for an individual with covariates $X^*$
and $Z^*$ is given by $\exp(\bar{\mu}+\bar{\tau}
X^*+\bar{\lambda}Z^*)$. The variability of
$\bar{\mu}$, $\bar{\tau}$, and $\bar{\lambda}$ may be estimated using
(\ref{mivar}), and confidence intervals may be computed based on the
$t$ distribution using (\ref{midf}), as described previously.

\section{Simulations}
\label{s:sims}
Data with missing failure indicators were simulated, and several
possible methods were
compared with respect to bias, coverage, and confidence interval width.
Survival times for 1,000 individuals were generated with exponentially
distributed failure times under a proportional hazards model with
covariates as proposed by Bender et al. \cite{bender}. That is, the
survival time for each individual was distributed according
to (\ref{coxprophaz}) where $\lambda_0(t)=1$ is the baseline hazard. 
For our simulations, $Z_i$ was a single baseline covariate following a normal
distribution with mean 2 and unit variance. In other words,
conditional of $Z_i$, the
failure times $T_i$ followed an exponential distribution with hazard
$\exp(\beta^\prime Z_i)$ where $\beta\in\{-0.5,-1.5,-3\}$. The
censoring times $C_i$ followed an exponential distribution with mean
5 (corresponding to a hazard of $\exp(-\log(5)) \approx
\exp(-1.61)$). This yielded about 35\%, 75\% and 90\% censoring for
$\beta=-0.5$, $\beta=-1.5$, and $\beta=-3$, respectively. We also
defined $\Delta_i=I(T_i \leq C_i)$. If $\Delta_i=0$, the implication
is that the follow up period ended before the participant developed
TMD, meaning that the observation was censored at time $C_i$.

Covariates are represented by $Z_i$, a risk factor for TMD measured
at enrollment, and $X_i$, a
measurement collected on the last screener. 
For each observation, a normally distributed covariate $X_{i1}$
was generated with mean $\Delta_i$ and standard deviation 0.3.  
In OPPERA, $X_i$ represents a question on the screener evaluating
some symptom of first-onset TMD, such as the frequency of jaw pain.  
This was used to generate $Q_i=I(X_i>0.5)$, an indicator of whether participant $i$
screened positive on their last screener. Note that $X_i$ depends
on $\Delta_i$, since participants who developed first-onset TMD are
more likely to report symptoms on their screener, and $Q_i$ depends on
$X_i$, since the screener is positive if enough symptoms are
reported.
%
Also, 
$\xi_i=I(\Delta_i\text{ is observed})$
corresponds to the indicator of whether participant $i$ came in for
their clinical exam if $Q_i=1$. In all simulations, $\delta_i$ was
used as the failure indicator rather than $\Delta_i$, where $\delta_i$
is defined as
\[
\delta_i=
\begin{cases}
\Delta_i &\text{if $Q_i=1$} \\
0 & \text{if $Q_i=0$}
\end{cases}
\]
In other words, we set the failure indicator $\delta_i=0$ if the final
screener was negative. This decision was made to reflect the fact that
OPPERA participants who had a negative screen were not examined. Hence
it is possible that some participants developed first-onset TMD but
were never examined due to their final screener being negative. Thus,
the simulations (incorrectly) treat these observations as censored.


We created missing failure indicators under the following 
classical missing data mechanisms of Rubin \cite{rubinmiss}:
\begin{enumerate}
\item The probability of having a missing failure indicator is
  independent of the data.  This is known as missing completely at
  random (MCAR).\label{MCAR}
\item The probability of having a missing failure indicator depends
  on an observed covariate. This is known as missing at random
  (MAR).\label{MAR}
\item The probability of having a missing failure indicator depends
 on the (potentially unobserved) failure indicator. 
This is known as missing not at random (MNAR).\label{MNAR}
\end{enumerate}

Our method assumes that the data are MAR, which includes MCAR as a
special case. Our simulations under MAR and MNAR parallel the study
protocol in that failure indicators can only be missing for those
with positive screeners.
In other words, observations were potentially missing if
and only if $Q_i=1$. (Individuals with negative screeners have $Q_i=0$ and are assumed to
be censored. Those with positive screeners have $Q_i=1$ and may have missing clinical examinations.)
Details and results for MCAR and MNAR data are shown in Sections
\ref{simmcar} and \ref{simmnar} in the Supporting Information. We also
considered several simulation scenarios where the logistic regression
model for predicting the failure indicator was misspecified; see
Section \ref{badlogregapp} in the Supporting Information.
For MAR data, we set failure indicators to be missing with
probability
\begin{equation} \label{missingdatmech}
P(\xi_i=0|X_i,Z_i,V_i,Q_i=1)=\frac{\exp(-0.2-0.3Z_i+0.1V_i)}{1+\exp(-0.2-0.3Z_i+0.1V_i)}
\end{equation}
This resulted in approximately 50\% of failure indicators being set
to missing, which is consistent with the rate of missing failure
indicators in the OPPERA study.

In each simulated data set, all observations with observed failure indicators
who had a positive screen were used to fit a
logistic regression model for case status with covariates $Z_i$, $X_i$ and $V_i$.
That is, using the complete data (i.e. observations with $Q_i=1$ and
$\xi_i=1$),
we fit the logistic
regression model for the event probability conditional
on $Z_i$, $X_i$, and $V_i$, namely
\begin{equation}\label{originalsim}
\logit\{P(\Delta_i=1|X_i,Z_i,V_i,Q_i=1,\xi_i=1)\}=\alpha^\prime
X_i+\gamma^\prime Z_i + \eta V_i
\end{equation}
The estimated probabilities
$\hat{p}_i=\frac{\exp(\hat{\alpha}^\prime X_i+\hat{\gamma}^\prime Z_i
  + \hat{\eta}V_i)}{1+\exp(\hat{\alpha}^\prime X_i+\hat{\gamma}^\prime
  Z_i+\hat{\eta}V_i)}$
were calculated for individuals with $Q_i=1$ (where $\hat{\alpha}$,
$\hat{\gamma}$, and $\hat{\eta}$ are drawn from their posterior distribution).

To evaluate the performance of our method, multiple imputation was employed
to calculate 10 imputed estimates of $\beta$ for each simulation as
described in
Section \ref{mi}. For each observation $i$
with $Q_i=1$ and $\xi_i=0$, we estimated failure indicators
$\hat{\Delta}_{ij}$ independently for each imputation $j$.  

A Cox proportional hazards model was fit for each imputed data set,
and the imputed estimates of the regression coefficient and their
variances were recorded.
These were aggregated using equations (\ref{mimean}) and (\ref{mivar})
to create confidence intervals for the multiple imputation estimates.

The performance of our method 
was compared with that of the method of Cook and Kosorok
\cite{kosorok}. 
To obtain the estimates of Cook and Kosorok \cite{kosorok},
for each simulated data set, we estimated the probabilities
$\hat{p}_i$ that the (potentially unobserved) event for participant $i$ is a
true event, as described previously. We then fit a weighted Cox
proportional hazards model to the data set with weights calculated as
follows: Each observation with a missing failure indicator was deleted
and replaced with two new observations.
Each such pair of observations had the same failure time and
covariates, but different failure indicators
and weights.
The first observation had weight $\hat{p}_i$ and $\hat{\Delta}_i=1$,
and the second observation had weight $1-\hat{p}_i$
and $\hat{\Delta}_i=0$.
Participants with fully observed data retained a single observation in
the data set with unit weight. The estimated
regression coefficient, $\hat{\beta}$ was recorded.

The variance of
this estimate was estimated by generating 1,000 bootstrap replicates of each
simulated data set and refitting the model for each bootstrap
replicate. A set of 1,000 subjects was selected at each bootstrap
iteration by sampling from the data with replacement. For each
bootstrap replicate, the estimated probability $\hat{p}_i^*$ that
participant $i$ is a true failure was calculated. These estimated
$\hat{p}_i^*$'s were used to calculate a bootstrap estimate
$\hat{\beta}^*$ of $\beta$ using a weighted Cox model as described in
the previous paragraph.
  The average parameter
estimate, $\bar{\hat{\beta}}$ and percentile
confidence intervals $(\beta_{0.025},\beta_{0.975})$ were all recorded, where
$\beta_{\theta}$ is the $\theta^{th}$ quantile
among the 1,000 bootstrap replicates. 

We also compared our method to the ideal situation in which the true
values of $\Delta_i$ were observed for all observations (note that
$\Delta_i$ was used instead of $\delta_i$ in this case), complete case
analysis (meaning that we exclude from the
data set all observations with missing failure indicators), and
two ad hoc methods in which we treat the missing indicators either all 
as censored or all as failures. 
Results under the assumption of MAR
are shown in Table \ref{mar}. We estimated the bias of each method by
calculating the mean difference between the estimated Cox regression
coefficient and the true coefficient over the 1000 simulations. We
also calculated the mean width of the confidence intervals produced by
each method over the 1000 simulations. Similarly, we calculated the
empirical coverage probability for the confidence intervals produced
by each method by dividing the number of times that the confidence
intervals contained the true value of the parameter by 1000. We also
report the Monte Carlo error for the coverage rate, which is the error
in the empirical coverage probability due to conducting only a finite
number of simulations (which would be $\sqrt{\alpha(1-\alpha)/n}$ for
$n$ simulations). Finally, the rate of missing information and the
average running time of each method was
computed. 


All calculations were performed using R versions 3.0.2 running on a
single core of a Dell C6100 server with a 2.93 GHz Intel
processor. The function ``mi.binary'' in the ``mi'' R package was used
to generate the imputed values of the missing failure
indicators. The functions ``boot'' and ``boot.ci'' in the ``boot'' R
package were used to calculate the bootstrap estimates of the standard
error of the Cook and Kosorok \cite{kosorok} method. The Cox
proportional hazard models were fit using the ``coxph'' function in
the ``survival'' R package. The code used to perform the simulations
(and analyze the OPPERA data) is available in the Supporting
Information.

\begin{table}
\caption{Simulation Results for MAR}
\begin{center}
\begin{tabular}{clrllllll}
\toprule
$\beta$\footnotemark[1]&Method&Bias&SE (Bias)&Width&SE (Width)&Coverage\footnotemark[2]&Running Time (s.)\\ \hline
\midrule
-0.5&Full Data&-0.0008&0.0005&0.1666&0.0004&0.962&0.008\\
&Complete Case&0.0033&0.0007&0.2152&0.0004&0.955&0.007\\
&Treat all as Censored&0.1058&0.0007&0.2127&0.0004&0.514&0.007\\
&Treat all as Failures&0.0018&0.0005&0.1699&0.0004&0.964&0.008\\
&Cook \& Kosorok&-0.0009&0.0005&0.1728&0.0004&0.959&22.0\\
&Multiple Imputation&-0.0003&0.0005&0.1721&0.0004&0.961&0.49\\ \hline
-1.5&Full Data&0.0047&0.0011&0.3176&0.0002&0.938&0.008\\
&Complete Case&-0.0558&0.0015&0.4317&0.0003&0.927&0.007\\
&Treat all as Censored&0.1241&0.0014&0.421&0.0003&0.767&0.007\\
&Treat all as Failures&0.0716&0.0011&0.3154&0.0002&0.841&0.007\\
&Cook \& Kosorok&0.0052&0.0011&0.3399&0.0003&0.942&17.50\\
&Multiple Imputation&0.0082&0.0011&0.3353&0.0002&0.942&0.40\\ \hline
-3&Full Data&-0.0294&0.0025&0.7606&0.0009&0.945&0.007\\
&Complete Case&-0.2044&0.0036&1.0855&0.0017&0.918&0.008\\
&Treat all as Censored&0.0988&0.0034&1.0413&0.0015&0.92&0.008\\
&Treat all as Failures&0.5914&0.0025&0.6293&0.0006&0.085&0.008\\
&Cook \& Kosorok&-0.0302&0.0029&0.9078&0.0017&0.94&17.33\\
&Multiple Imputation&-0.0042&0.0028&0.8556&0.0014&0.947&0.43\\
\bottomrule
\hline \\
\end{tabular}
\end{center}
{\footnotesize \footnotemark[1]: The rate of missing information is
  $0.017$ when $\beta=-0.5$, $0.061$ when $\beta=-1.5$, and $0.178$
  when $\beta=-3$.} \\
{\footnotesize \footnotemark[2]: The Monte Carlo error is 0.007.}
\label{mar}
\end{table}

The empirical coverage probability of the confidence intervals
produced by multiple imputation is close to the nominal level (0.95)
in all simulations.
Our multiple imputation method and the method of Cook and Kosorok
\cite{kosorok}
produced approximately unbiased estimates and valid confidence
intervals in all the
scenarios we considered. The estimates produced by the other methods
showed a larger amount of bias and did not always achieve
the desired coverage level. Our multiple
imputation method also yielded the narrowest confidence intervals in
each scenario. Although the method of Cook and Kosorok \cite{kosorok}
produced confidence intervals that were only slightly wider, this
indicates that our proposed method may have slightly greater power to
detect true associations, particularly when the absolute value of
$\beta$ is large. Our proposed method also tended to have lower bias
than the method of Cook and Kosorok \cite{kosorok} when the absolute
value of $\beta$ is large. The running time of our
proposed method was also significantly less than the running time of
the Cook and Kosorok \cite{kosorok} method. 
Moreover, for most parameter values, the coverage
probabilities for the complete case and ad hoc methods were significantly different
($p<0.01$) from the nominal rate.

In addition, we examined the performance of our proposed methods when
we changed the logistic regression model for $\Delta_i$.  We investigate two
additional types of models: one in which the model contained a variable unrelated to
case status and another in which the model does not include one
variable related to case status.
As in the previous simulations,
the failure times were generated by (\ref{coxprophaz}), censoring was exponential
with mean 5, failure indicators were set to be missing completely at random or
missing at random with probability given in equation (\ref{missingdatmech}),
$Z_i\sim N(2,1)$, $X_{i1}\sim N(\Delta_i,0.3)$ and
$Q_i=I(Y_{i2}>0.5)$
for $i=1,\ldots,n$. 
We also generated $X_{i2}\sim N(0,1)$ where $Z_i,X_{i1},X_{i2}$ were
mutually independent and $X_{i2}$ was independent of $\Delta_i$ and $Q_i$.

In the previous simulations, we fit the data to (\ref{originalsim})
with covariates $Z_i$ and $X_i=X_{i1}$. 
The additional simulations instead used the covariates and parameters as follows:
\begin{enumerate}
\item $\tilde{X}_i=\{1, X_{i1}, X_{i2}\}$ \label{toobig}
\item $\tilde{X}_i=0$ \label{toosmall}.
\end{enumerate}

That is, rather than fitting model (\ref{originalsim}) to the data,
we modeled the case probability with
\begin{equation}
\logit\{P(\Delta_i=1|X_i,Z_i,V_i,Q_i=1)\}=\tilde{\alpha}^\prime\tilde{X}_i+\gamma
Z_i+\eta V_i.
\end{equation}
The results, which are shown in Section \ref{badlogregapp} in the
Supporting Information,
remained similar under both alternative models.
This indicates that the proposed methods are
robust to misspecification of the logistic regression model in some
situations. Most notably, leaving out one covariate
that was weakly related to case status did not markedly decrease the
performance of the method.

We also performed some simulations where a random subset of the
observations with $Q_i=0$ were set to have missing failure
indicators. The model to predict $\Delta_i$ was fitted using only the
observations for which $Q_i=1$, but the model was applied to all
observations with missing failure indicators (including observations
where $Q_i=0$). The results are shown in Section \ref{badlogregapp}
in the Supporting Information. In this case our method (as well as the
Cook and Kosorok \cite{kosorok} method) produced reasonable results
when the logistic regression model was specified correctly or when an
extra covariate was included in the model. However, both methods
performed poorly when an important covariate was missing from the
logistic regression model.

Finally, we conducted simulations to evaluate the method's ability to
estimate incidence rates.
A similar multiple imputation strategy was applied to Poisson regression.
Our method produced estimates much
closer to the true incidence rates than the complete case
estimate. In fact, the complete case method underestimated
incidence rates by as much as a factor of 3.
See Section \ref{simpois} in the Supporting Information for details.

\section{Analysis of the OPPERA Study}
\label{s:inf}
In this section, we apply our method to estimate hazard ratios and
incidence rates in the OPPERA study using $m=10$ imputations.

\subsection{Hazard Ratios}
We applied our method to the OPPERA cohort to adjust for the effect of
participants with missing clinical examinations. (Note that examinations for participants evaluated by
Examiner \#4 were also treated as missing.) First, we estimated
the probability that a participant would be diagnosed as an
incident case of TMD given a positive screener. Due to the
rich body of information collected in each screener, we carefully selected
a small number of predictor variables. Specifically, we fit a
logistic regression model to predict the result of the clinical exam
based on each item in the screener. As described previously, the
regression coefficients were assumed to have a prior distribution that
was Cauchy with center 0 and scale 2.5. All models were adjusted for
study site.

The majority of
the variables measured on the screener were not associated with the result
of the clinical examination. The strongest predictor of being
diagnosed with TMD was a
count of non-specific orofacial symptoms (e.g stiffness, fatigue) in
the previous three months.
The time elapsed since enrollment and OPPERA study site were also important
covariates, as shown in Bair et al. \cite{bairOpperaII}.
Several other possible predictors of
being diagnosed with TMD were identified, but including these additional
predictors in the model did not improve the predictive accuracy of
the model and hence they were not included. (In general failure to
include a relevant predictor variable when performing multiple
imputation will produce greater error than including an irrelevant
variable as evidenced by our simulations, so generally it is better to
err on the side of including too many predictors rather than too
few. However, in this case, our testing indicated that included
additional variables did not improve the predictive accuracy of the
model and in fact might actually decrease the accuracy. Hence, in this
case we favored the more parsimonious model.)

Thus, we estimated the probability of being diagnosed with TMD based
on the count of non-specific orofacial symptoms,
time since enrollment, and OPPERA study site. This model was used to perform
multiple imputation for those with no clinical examination. These
imputed data sets were used to fit a series of Cox proportional
hazards models to estimate the hazard ratio (and associated confidence
interval and p-value) for each predictor using the methods described in
section \ref{mi}. Examples of predictors include perceived stress,
history of comorbid chronic pain conditions, and smoking status.

In addition, Bair et al. \cite{bairOpperaII} examined univariate relationships
between examination attendance and numerous possible predictor
variables. Differences between examined and non-examined participants
were small and most were not statistically significant.
However, a few of the differences were statistically significant,
indicating that the data were not MCAR, since MCAR requires that the
probability of a missing observation does not depend on the data.

Table \ref{OPPERAresults} shows the results of applying our method to
a subset of the putative risk factors of TMD measured in OPPERA. Due
to the large number of putative risk factors
measured in OPPERA, we
only report the results for a selected subset of the variables.
All continuous variables were normalized to have
mean 0 and standard deviation 1 prior to fitting the Cox
models. (Thus, the hazard ratios for the continuous variables
represent the hazard ratios corresponding to a one-standard deviation
increase in the predictor variable.)
In Table \ref{OPPERAresults}, all
the quantitative sensory testing and psychosocial variables were
continuous, while all of the clinical variables were dichotomous (and
hence were not normalized). The small number of missing values in
these predictor variables were (singly) imputed using the EM
algorithm; see Greenspan et al. \cite{GreenspanQST} or Fillingim et
al. \cite{FillingimPsych} for details. For a more detailed description
of the OPPERA domains, see Section \ref{opperaappendix} in the
Supporting Information, Maixner et al. \cite{maixner}, and Slade et
al. \cite{slade}.

\begin{table}[h!]
\centering
\caption{Results from the OPPERA Study}
\label{OPPERAresults}
{\small 
    \begin{tabular}{lrrrrrrrr}
\toprule
\multicolumn{5}{c}{\textbf{Treat All MCIs as Censored}} &
\multicolumn{4}{c}{\textbf{Multiple Imputation}}     \\ \hline
& HR    & LCL & UCL & P     &       HR    & LCL & UCL & P\\ \hline
\midrule
    \multicolumn{9}{l}{Clinical Variable}   \\  \hline
    In the last month & \multirow{2}{*}{3.26}  & \multirow{2}{*}{1.83}  & \multirow{2}{*}{5.84}  & \multirow{2}{*}{$<$0.0001} & \multirow{2}{*}{2.35}  & \multirow{2}{*}{1.39}  & \multirow{2}{*}{3.96}  & \multirow{2}{*}{0.0015} \\
    could not open mouth wide &&&&&&&& \\ 
Has two or more comorbid& 3.08  & 2.26  & 4.21  & $<$0.0001 & 2.36  & 1.79  & 3.11  & $<$0.0001 \\
chronic pain disorders&&&&&&&& \\ 
         History of 5 respiratory conditions & 1.38  & 1.01  & 1.87  & 0.0408  & 1.44  & 1.13  & 1.85  & 0.0040 \\
    Smoking: current & 1.26  & 0.86  & 1.84  & 0.2403  & 1.48  & 1.07  & 2.04  & 0.0166 \\
    Smoking: former & 1.87  & 1.22  & 2.87  & 0.0041 & 1.70  & 1.18  & 2.46  & 0.0045 \\
    One or more palpation tender& \multirow{2}{*}{1.83}  & \multirow{2}{*}{1.32}  & \multirow{2}{*}{2.52}  & \multirow{2}{*}{0.0002} & \multirow{2}{*}{1.54}  & \multirow{2}{*}{1.18}  & \multirow{2}{*}{2.02}  & \multirow{2}{*}{0.0018} \\
 points: right temporalis &&&&&&&& \\ 
     One or more palpation tender & \multirow{2}{*}{1.60}  & \multirow{2}{*}{1.14}  & \multirow{2}{*}{2.25}  & \multirow{2}{*}{0.0064} & \multirow{2}{*}{1.50}  & \multirow{2}{*}{1.13}  & \multirow{2}{*}{1.98}  & \multirow{2}{*}{0.0045} \\
points: left temporalis      &&&&&&&& \\ 
     One or more palpation tender & \multirow{2}{*}{1.85}  & \multirow{2}{*}{1.35}  & \multirow{2}{*}{2.53}  & \multirow{2}{*}{0.0001} & \multirow{2}{*}{1.69}  & \multirow{2}{*}{1.31}  & \multirow{2}{*}{2.17}  & \multirow{2}{*}{$<$0.0001} \\
points: right masseter &&&&&&&& \\ 
     One or more palpation tender & \multirow{2}{*}{1.70}  & \multirow{2}{*}{1.23}  & \multirow{2}{*}{2.35}  & \multirow{2}{*}{0.0013} & \multirow{2}{*}{1.50}  & \multirow{2}{*}{1.15}  & \multirow{2}{*}{1.97}  & \multirow{2}{*}{0.0031} \\  
points: left masseter &&&&&&&& \\ \hline         
    \multicolumn{9}{l}{Quantitative Sensory Testing Variable}                                                              \\ \hline
    Pressure pain threshold: temporalis & 1.26  & 1.07  & 1.49  & 0.0065 & 1.14  & 1.00  & 1.31  & 0.0466 \\
    Pressure pain threshold: masseter & 1.23  & 1.04  & 1.45  & 0.0170 & 1.14  & 0.99  & 1.31  & 0.0674 \\
    Pressure pain threshold: TM joint & 1.25  & 1.05  & 1.48  & 0.0106 & 1.15  & 1.01  & 1.32  & 0.0416 \\
    Mechanical pain aftersensation: & \multirow{2}{*}{1.23}  & \multirow{2}{*}{1.09}  & \multirow{2}{*}{1.38}  & \multirow{2}{*}{0.0006} & \multirow{2}{*}{1.15}  & \multirow{2}{*}{1.04}  & \multirow{2}{*}{1.28}  & \multirow{2}{*}{0.0071} \\
512mN probe, 15 s &&&&&&&& \\
    Mechanical pain aftersensation: & \multirow{2}{*}{1.20}  & \multirow{2}{*}{1.07}  & \multirow{2}{*}{1.34}  & \multirow{2}{*}{0.0020}  & \multirow{2}{*}{1.12}  & \multirow{2}{*}{1.02}  & \multirow{2}{*}{1.24}  & \multirow{2}{*}{0.0241} \\
    512mN probe, 30 s &&&&&&&& \\ \hline
    \multicolumn{9}{l}{Psychosocial Variable}                                                            \\ \hline
    PILL Global Score
    & 1.52  & 1.35  & 1.71  & $<$0.0001 & 1.42  & 1.29  & 1.58  & $<$0.0001 \\
    EPQ-R Neuroticism & 1.39  & 1.21  & 1.60  & $<$0.0001 & 1.25  & 1.11  & 1.42  & 0.0003 \\
    Trait Anxiety Inventory & 1.43  & 1.25  & 1.64  & $<$0.0001 & 1.34  & 1.19  & 1.52  & $<$0.0001 \\
    Perceived Stress Scale & 1.35  & 1.17  & 1.55  & $<$0.0001 & 1.29  & 1.15  & 1.44  & $<$0.0001 \\
    SCL 90R Somatization & 1.44  & 1.31  & 1.58  & $<$0.0001 & 1.40  &
    1.29  & 1.51  & $<$0.0001 \\  \hline
\bottomrule
    \end{tabular}%
}
\end{table}

The rate of missing information varied slightly for each putative risk
factor. The average rate of missing information was approximately
$0.097$. Compared to the unimputed results, which treated missing
failure indicators as censored observations, imputation slightly
reduced the hazard ratios for most of the psychosocial variables that
were measured in OPPERA. For instance, Table \ref{OPPERAresults} shows
the (standardized) hazard ratios for the Pennebaker Inventory of
Limbic Languidness (PILL) score, the neuroticism subscale of the
Eysenck Personality Questionnaire (EPQ), the Spielberger Trait Anxiety
Inventory score, the Perceived Stress Scale, and the somatization
subscale of the Symptom Checklist-90, Revised (SCL-90R).
In each case, the hazard ratios were reduced after imputation.

A similar pattern was observed after applying our imputation method to
the measures of experimental pain sensitivity. The mechanical pain aftersensation ratings were
strongly associated with first-onset TMD before imputation, but they
were only weakly associated with first-onset TMD after imputation. The
pressure pain algometer ratings were also more weakly associated with
TMD after imputation (and two of three ratings in Table
\ref{OPPERAresults} were no longer significantly associated with
first-onset TMD at the $p<0.05$ level). 

Interestingly, the hazard ratios for the presence of one or more
palpation tender points at the temporalis and masseter muscles were also
attenuated after imputation. 
These tender points were evaluated as part of the clinical examination using
a different protocol than the quantitative sensory testing
algometer pain ratings. However, both pain measures
(algometer and palpation) were measured at the same facial 
locations. 
While the palpation ratings were more
strongly associated with first-onset TMD than the algometer ratings
both before and after imputation, it is interesting that different
pain sensitivity measures using different protocols at the same
anatomical location were both attenuated by imputation.

The effects of other clinical variables were also attenuated after
imputation. For example, the hazard ratios associated with being
unable to open one's mouth wide in the past month and having two or
more comorbid pain conditions were both noticeably attenuated after
imputation. However, other clinical variables were more strongly
associated with first-onset TMD after imputation. For example, having
a history of respiratory illness was only
weakly associated with first-onset TMD before imputation (HR=1.38,
p=0.04), but the association was much stronger after imputation
(HR=1.43, p=0.004). Also, being a current smoker was not significantly
associated with first-onset TMD before imputation (HR=1.26, p=0.24)
but was associated after imputation (HR=1.49, p=0.02). 

\subsection{Incidence Rates}
In Table \ref{tmdrates}, the incidence rate of first-onset TMD was
estimated using two different approaches.
First, all missing failure indicators were treated as censored. Second,
the multiple imputation method in this paper was used to estimate the
incidence rate.
The estimated TMD incidence rate using multiple imputation was 66\%
greater than the unimputed estimate. The estimated incidence rate increased
by 70\% for females and 87\% for males. Estimated incidence rates for
whites and Hispanics were 118\% and 202\% higher, respectively, with
imputation. Thus, the incidence rate is likely to be
underestimated without imputation.
\begin{center}
\begin{table}[h!]
\caption{Estimated TMD Incidence Rates With and Without Imputation}
\label{tmdrates}
\begin{center}
\begin{tabular}{lrrr}
\toprule
&No MI&	MI&	Percent Change\\ \hline
\midrule
Overall	&2.23&	3.78&	70\%\\ \hline
Males	&1.87	&3.49	&87\%\\ 
Females&	2.46	&4.19&	70\%\\ \hline
White&	1.70	&3.70	&118\%\\
Black	&4.20	&5.70	&36\%\\
Hispanic	&1.17	&3.53	&202\%\\
Other	&1.10&	1.86&	69\%\\
\bottomrule
\hline \\
\end{tabular}
\end{center}
{\footnotesize
Incidence rates are given in cases per 100 person-years.}
\end{table}
\end{center}
\section{Discussion}
\label{s:discuss}
We have developed a computationally efficient method
to adjust for missing failure indicators in time-to-event data using
logistic regression and multiple imputation.
Logistic regression  is used to estimate the failure probability for
participants with missing failure indicators.  The missing values are
imputed, and the standard errors are estimated using our multiple imputation method.
This framework is important in studies where failure
status may be measured in stages, which may lead to missing failure
status indicators. This is a common occurrence in
studies of diseases that are difficult or expensive to diagnose, such
as TMD.

The present method is similar to the method of Magder and Hughes
\cite{MagderHughes},
who use an iterative procedure for
parameter estimation based on the EM algorithm. Our assumption of MAR data renders their iterative
method unnecessary. 
%
%
Other methods \cite{mckeaguesubram, gijbels, Sundarraman2000}
depend on the MCAR assumption, which does not hold for the OPPERA study.
Chen et al. \cite{chenheshensun} estimate Cox regression parameters
using the EM algorithm and establish their consistency under basic
regularity conditions, including missing at random (MAR) failure indicators.
However, their approach depends on the assumptions of
piecewise constant proportional hazard functions for
the censoring time as well as for the failure time.

In each simulation scenario, our multiple imputation method produced the
narrowest valid confidence intervals and no
significant bias. In particular, the
method of Cook and Kosorok \cite{kosorok}
produced slightly wider confidence intervals
in all but one of the simulations we considered. 
The differences were
small, so the performance of the two methods appear to be
comparable for most practical purposes. However, we believe that our
method 
has several possible advantages over the method of
Cook and Kosorok \cite{kosorok}.
First, 
bootstrapping is much more intensive
computationally than our multiple imputation approach. Calculating
bootstrap confidence intervals generally requires at least 1000
bootstrap replicates \cite{ET93}, whereas as few as 10 imputed data
sets may be sufficient for multiple imputation
\cite{rubinlittle}. Although the difference in the computing time of
the two methods is small for a single fitted model, many such models
will be required in the course of the OPPERA study. OPPERA has
already collected data on approximately 3000 genetic
markers and has plans to collect data on approximately a million
genetic markers in a genome-wide association study. Thus, at least
3000 (and potentially as many as a million) Cox models will need to be
fit, and our proposed method may allow for a significant decrease in
computing time.
Moreover, our method can also be easily implemented in
popular statistical software packages (such as SAS) without additional
programming.

Additionally, our methodology may easily be extended to other 
models, such as Poisson regression.  We conducted simulations
(Table \ref{poissonmar} in the Supporting Information) that showed
that our
proposed method can be used to estimate incidence rates using Poisson
regression, which is one of the research aims of the OPPERA study.
In particular, estimates of the failure rates were biased when
missing failure indicators were treated as censored
or when the complete case method was used, but they were unbiased
when we employed the methodology in this paper.

Our method may yield increased bias and decreased coverage
if the logistic regression model for predicting case status is
inaccurate, as observed in the simulations in Section \ref{badlogregapp}
in the Supporting Information.
However, this would also be true for competing methods, including the
method of Cook and Kosorok \cite{kosorok}.


Our proposed also requires that the missing data be MAR. Although it
is impossible to test this assumption directly, Bair et
al. \cite{bairOpperaII} showed that there were no significant
differences between those who did and not attend their clinical
examination with respect to a wide range of demographic variables and
putative risk factors for TMD. Thus, the MAR assumption is reasonable
for OPPERA. Furthermore, the results of the simulations described in
Section \ref{simmnar} show that our proposed method can produce valid
results in some situations even if the MAR assumption is violated.

Also, our proposed method is only useful for imputing missing event
failure indicators among participants who have positive
screeners. If a participant develops first-onset TMD but still has a
negative screener, such a participant will be treated as censored, and
our method is unable to correct for this misclassification. The OPPERA
screener was designed to have high sensitivity and modest specificity,
so the number of false negative screens is expected to be
low. (Indeed, OPPERA performed clinical examinations on a subset of
the participants with negative screeners. Although analysis of this
data is ongoing, preliminary results suggests that the false negative
rate is less than 5\%.) Thus, we expect that the small number of false
negative screens will not meaningfully affect the results of our
analysis. Also, note that under our simulation scenarios, we assumed
that some failures were not observed due to a negative screener. Since
our proposed method gave satisfactory results in these simulation
scenarios, it appears that failing to observe some events due to
negative screeners should not significantly bias the results.

In the OPPERA study, the hazard ratios associated with some variables
were noticeably different after imputation.
Although other results remained qualitatively
unchanged, we
note that even small changes in hazard ratios are important.
In addition, estimated incidence rates were significantly increased after
imputation. Since the results of OPPERA may become normative in the
orofacial pain literature, precise calculation of the incidence rate
of TMD and the hazard ratios associated with putative risk factors is
important. Thus, imputation is recommended. 

\acks
The authors would like to acknowledge and thank the principal
investigators of the OPPERA study, namely William Maixner, Luda
Diatchenko, Bruce Weir, Richard Ohrbach, Roger Fillingim, Joel
Greenspan, and Ronald Dubner. The OPPERA study was supported by
NIH/NIDCR grant U01DE017018. Naomi Brownstein was supported by
NIH/NIEHS T32ES007018 and NSF 
Graduate Research Fellowship Program grant 
0646083. Jianwen Cai was supported by NIH/NCI grant P01CA142538 and
NIH/NIEHS grant R01ES021900. Eric Bair was supported by NIH/NIDCR
grant R03DE023592, NIH/NCATS grant UL1TR001111, and NIH/NIEHS grant
P03ES010126.

\bibliographystyle{wileyj}
\bibliography{bibliography}

\newpage

\let\cleardoublepage\clearpage
\setcounter{page}{1}
\renewcommand{\thesection}{S\arabic{section}}
\setcounter{section}{0}
\renewcommand{\thefigure}{S\arabic{figure}}
\setcounter{figure}{0}
\renewcommand{\thetable}{S\arabic{table}}
\setcounter{table}{0}
\renewcommand{\theequation}{S.\arabic{equation}}
\setcounter{equation}{0}
\noindent
{\bf \LARGE Web-based Supplementary Materials for ``Parameter Estimation in Cox 
Models with Missing Failure Indicators'' by Naomi C Brownstein,
Jianwen Cai, Gary Slade, and Eric Bair}

\section{Description of the OPPERA Study}\label{opperaappendix}

The primary objective of the OPPERA study is to identify possible risk
factors for developing first-onset TMD. See Maixner et al. \cite{maixner},
Slade et al. \cite{slade}, and Bair et al. \cite{bairOpperaII} for a
more detailed description of the study.
The risk factors considered in OPPERA are classified into the
following domains: sociodemographic, clinical, psychosocial,
autonomic,  quantitative sensory testing (QST), and genetics.
The remainder of this section describes these OPPERA domains in more
detail.

First, sociodemographic information was recorded for each OPPERA
participant. This includes age, gender, race, and OPPERA study site, as well
as educational attainment, income, and marital status.  For example,
TMD is more common in females than males and in non-Hispanic whites
than in other races. Details are provided in Slade et al. \cite{slade}.

Clinical risk factors refer to variables that ``typically are
considered in clinical settings when evaluating patients''
\cite{OhrbachClinical}. These clinical variables may be evaluated via
physical examinations or
questionnaires. Examples include  headaches, back
aches, pain in other regions of the body, jaw mobility, jaw noises, and
orofacial trauma. OPPERA participants also self-reported their health
history, including the presence of comorbid pain conditions such as
irritable bowel syndrome, fibromyalgia, and dysmenorrhea.

Psychosocial factors have also been shown to be associated with TMD
\cite{FillingimPsych}. Specific qualities related to psychosocial
functioning were evaluated in OPPERA, including general psychological
function, affective distress, psychological stress, somatic awareness,
and coping/catastrophizing. Affective distress measures include state
and trait anxiety and mood. Psychological stress includes
perceived stress and measures of post-traumatic stress
disorder. Somatic awareness assesses sensitivity to
physical sensations. Finally, coping/catastrophizing assesses
individuals' ability to handle pain.

The association between TMD and the function of the autonomic nervous
system was also evaluated. Key measures of autonomic function include
blood pressure, heart rate, and heart rate variability, which were
measured during the OPPERA baseline medical examination. In
previous studies, TMD was associated with higher heart rates and lower
heart rate variability, which are symptoms of dysregulation of the
autonomic nervous system. See Maixner et al. \cite{MaixnerAutonomic}
for a more detailed description of the autonomic data collected in
OPPERA.

The QST variables collected in OPPERA measure sensitivity to
experimental pain. Several measures of experimental pain sensitivity
were collected, including pressure pain thresholds measured by
algometers, mechanical (pinprick) pain sensitivity, and thermal pain
sensitivity.  See Greenspan et al. \cite{GreenspanQST} for a more
detailed description of these QST variables.

Finally, the association between TMD and selected genetic markers
was evaluated. A total of 3295 single nucleotide polymorphisms (SNP's)
were selected from genes that are believed to be associated with
pain. See Smith et al. \cite{SmithGenetics} for more detail on how the
SNP's were chosen and their association with TMD.

\section{Results of Additional
  Simulations}\label{moresimsappendix}
\subsection{Overview of Additional Simulations}
In this appendix, we provide the results of additional simulations.
We investigate the performance of the method under a variety of missing
data mechanisms. We also consider scenarios where the logistic
regression model for estimating the probability of being a case is
misspecified.

Recall that we created missing failure indicators under the following 
classical missing data mechanisms of Rubin (1976):
\begin{enumerate}
\item The probability of having a missing failure indicator is
  independent of the data.  This is known as missing completely at
  random (MCAR).\label{MCARappend}
\item The probability of having a missing failure indicator depends
  on an observed covariate. This is known as missing at random
  (MAR).\label{MARappend}
\item The probability of having a missing failure indicator depends
 on the failure indicator itself. This is known as missing not at
 random (MNAR).\label{MNARappend}
\end{enumerate}
%

\subsection{Additional Simulations Under MCAR} \label{simmcar}
In order to more closely parallel the OPPERA study, we simulated data
where we randomly set 40\% of the failure indicators to be missing
for those with $Q_i=1$. (Note that our simulations assume that
failure indicators can only be missing when $Q_i=1$. Without this
assumption the data would not be MCAR in this scenario, since $Q_i$
depends on $X_i$, which is observed.) 
This setup assumes that the probability that a
participant has a non-missing failure indicator depends only on whether or not
their screener was positive. The logistic regression model in this case included the
covariates $Z_i$ and $X_i$ as before, as well as the time of the
screener.
Results are shown in Table \ref{marbutmcarlike}.
All methods had a negligible amount of bias in these scenarios except for
the complete case method and
the method that treated all missing indicators as failures.  In
these simulations, the complete case method also displayed extreme
bias and poor coverage.  This indicates that a complete case
analysis would not be appropriate for a study such as OPPERA.

\renewcommand{\tablename}{Table}
\begin{table}
\caption{Simulation Results for MCAR}
\begin{center}
\begin{tabular}{clrcccll}
\toprule
$\beta$\footnotemark[1]&Method&Bias&SE (Bias)&Width&SE (Width)&Coverage\footnotemark[2]&Running Time (s.)\\ \hline
\midrule
-0.5&Full Data&0.0003&0.0005&0.1667&0.0001&0.95&0.008\\
&Complete Case&-0.0518&0.0007&0.2203&0.0001&0.854&0.007\\
&Treat all as Censored&0.0005&0.0007&0.2205&0.0001&0.952&0.008\\
&Treat all as Failures&0.0056&0.0006&0.1699&0.0001&0.953&0.007\\
&Cook \& Kosorok&0&0.0006&0.1746&0.0001&0.955&22.36\\
&Multiple Imputation&0.0002&0.0006&0.173&0.0001&0.958&0.51\\ \hline
-1.5&Full Data&-0.0018&0.0011&0.3184&0.0002&0.942&0.008\\
&Complete Case&-0.1303&0.0014&0.4257&0.0003&0.798&0.007\\
&Treat all as Censored&-0.0063&0.0014&0.4218&0.0003&0.954&0.007\\
&Treat all as Failures&0.0817&0.0011&0.3149&0.0002&0.808&0.007\\
&Cook \& Kosorok&-0.0021&0.0011&0.3426&0.0004&0.943&17.64\\
&Multiple Imputation&0.001&0.0011&0.3395&0.0002&0.944&0.39\\ \hline
-3&Full Data&-0.0152&0.0025&0.7561&0.0008&0.953&0.007\\
&Complete Case&-0.2332&0.0035&1.0332&0.0015&0.894&0.007\\
&Treat all as Censored&-0.0191&0.0033&1.0066&0.0014&0.959&0.008\\
&Treat all as Failures&0.6654&0.0024&0.6166&0.0006&0.047&0.007\\
&Cook \& Kosorok&-0.0186&0.0028&0.8938&0.0016&0.939&16.38\\
&Multiple Imputation&0.0073&0.0027&0.8493&0.0014&0.958&0.4\\
\bottomrule
\hline\\
\end{tabular}
\end{center}
{\footnotesize \footnotemark[1]: The rate of missing information is
  $0.024$ when $\beta=-0.5$, $0.079$ when $\beta=-1.5$, and $0.19$
  when $\beta=-3$.} \\
{\footnotesize \footnotemark[2]: The Monte Carlo error is 0.007.}
\label{marbutmcarlike}
\end{table}

\subsection{Alternative Logistic Regression Models}\label{badlogregapp}
We considered several scenarios where the logistic regression model
for the probability of being a case is misspecified. Recall that we
originally modeled the probability of being a case as
\begin{equation}\label{logisticmodelapp}
P(\Delta_i=1|X_i,Z_i,V_i)=\frac{\exp(\alpha^\prime X_i+\gamma^\prime
  Z_i+ \eta V_i)}{1+\exp(\alpha^\prime X_i+\gamma^\prime Z_i+\eta V_i)}
\end{equation}

The original logistic model had the covariates
$Z_i$, $X_i$, and $V_i$ where
$Z_i\sim N(2,1)$ and $X_i\sim N(\Delta_i,0.3)$
are mutually independent for $j=1,2,3$
and $i=1,\ldots,n$.

Two alternative models were examined:
\begin{enumerate}
\item \label{toobigapp}
The first alternative model was of the
form (\ref{logisticmodelapp}) but used the covariates
$\tilde{X}_i=\{X_{i1}, X_{i2}\}$ and $V_i$
where $X_{i2}\sim N(0,1)$. This scenario was to used to evaluate the
robustness of the method when an extraneous covariate is included in
the model. (Note that $X_{i2}$ is independent of $Z_i$, $X_{i1}$,
$\Delta_i$, and $Q_i$.)
\item \label{toosmallapp}
The second alternative model was generated according to
(\ref{logisticmodelapp}) but
was fit with the covariates
$Z_i$ and $V_i$. 
In the context of OPPERA, this represents the scenario in which we
failed to include a covariate that is associated with first-onset
TMD.
\end{enumerate}
Tables \ref{toobigmar} and \ref{toosmallmar} indicate that our method
produces valid results even if a noisy variable is added to the model
or if an important variable is not included in the model.

\renewcommand{\tablename}{Table}
\begin{table}
\caption{Results for an Extra Covariate Included
in the Logistic Regression Model}
\begin{center}
\begin{tabular}{clrcccll}
\toprule
$\beta$\footnotemark[1]&Method&Bias&SE (Bias)&Width&SE (Width)&Coverage\footnotemark[2]&Running Time (s.)\\ \hline
\midrule
-0.5&Full Data&0.0016&0.0006&0.1667&0.0001&0.942&0.008\\
&Complete Case&0.0037&0.0007&0.2156&0.0001&0.938&0.008\\
&Treat all as Censored&0.1054&0.0007&0.213&0.0001&0.509&0.008\\
&Treat all as Failures&0.004&0.0006&0.1701&0.0001&0.952&0.008\\
&Cook \& Kosorok&0.0012&0.0006&0.174&0.0001&0.955&22.91\\
&Multiple Imputation&0.0018&0.0006&0.1724&0.0001&0.955&0.52\\ \hline
-1.5&Full Data&-0.0058&0.001&0.3185&0.0002&0.955&0.008\\
&Complete Case&-0.0693&0.0015&0.4332&0.0004&0.909&0.007\\
&Treat all as Censored&0.1139&0.0014&0.4222&0.0003&0.799&0.007\\
&Treat all as Failures&0.0622&0.001&0.3161&0.0002&0.877&0.007\\
&Cook \& Kosorok&-0.0056&0.0011&0.3408&0.0003&0.946&18.09\\
&Multiple Imputation&-0.0024&0.0011&0.3368&0.0002&0.953&0.4\\ \hline
-3&Full Data&-0.0242&0.0025&0.7606&0.0008&0.962&0.007\\
&Complete Case&-0.1941&0.0037&1.0845&0.0017&0.924&0.007\\
&Treat all as Censored&0.1045&0.0036&1.0405&0.0015&0.906&0.007\\
&Treat all as Failures&0.5995&0.0025&0.6291&0.0006&0.089&0.007\\
&Cook \& Kosorok&-0.0172&0.0028&0.909&0.0017&0.953&16.93\\
&Multiple Imputation&0.0111&0.0027&0.8578&0.0014&0.958&0.42\\
\bottomrule
\hline\\
\end{tabular}
\end{center}
{\footnotesize \footnotemark[1]: The rate of missing information is
  $0.017$ when $\beta=-0.5$, $0.063$ when $\beta=-1.5$, and $0.188$
  when $\beta=-3$.} \\
{\footnotesize \footnotemark[2]: The Monte Carlo error is 0.007.}
\label{toobigmar}
\end{table}

\renewcommand{\tablename}{Table}
\begin{table}
\caption{Results when a Relevant Covariate is Omitted from the Logistic Regression Model}
\begin{center}
\begin{tabular}{clrcccll}
\toprule
$\beta$\footnotemark[1]&Method&Bias&SE (Bias)&Width&SE (Width)&Coverage\footnotemark[2]&Running Time (s.)\\ \hline
\midrule
-0.5&Full Data&-0.0007&0.0006&0.1666&0.0001&0.939&0.007\\
&Complete Case&0.0046&0.0007&0.2153&0.0001&0.948&0.007\\
&Treat all as Censored&0.1062&0.0007&0.2128&0.0001&0.497&0.007\\
&Treat all as Failures&0.0021&0.0006&0.17&0.0001&0.942&0.007\\
&Cook \& Kosorok&-0.0006&0.0006&0.1738&0.0001&0.945&17.84\\
&Multiple Imputation&-0.0003&0.0006&0.1726&0.0001&0.947&0.36\\ \hline
-1.5&Full Data&0.0012&0.0011&0.3174&0.0002&0.951&0.007\\
&Complete Case&-0.0581&0.0015&0.4308&0.0004&0.908&0.006\\
&Treat all as Censored&0.1237&0.0014&0.4198&0.0003&0.761&0.007\\
&Treat all as Failures&0.0697&0.001&0.315&0.0002&0.844&0.007\\
&Cook \& Kosorok&0.0029&0.0011&0.3449&0.0004&0.942&14.54\\
&Multiple Imputation&0.0055&0.0011&0.3411&0.0003&0.939&0.3\\ \hline
-3&Full Data&-0.0146&0.0025&0.7562&0.0008&0.952&0.007\\
&Complete Case&-0.1925&0.0038&1.0793&0.0017&0.911&0.007\\
&Treat all as Censored&0.1076&0.0035&1.0352&0.0015&0.899&0.007\\
&Treat all as Failures&0.6021&0.0025&0.6261&0.0006&0.097&0.007\\
&Cook \& Kosorok&-0.0181&0.0029&0.9025&0.0017&0.935&14.11\\
&Multiple Imputation&0.0074&0.0029&0.8761&0.0015&0.936&0.32\\

\bottomrule
\hline\\
\end{tabular}
\end{center}
{\footnotesize \footnotemark[1]: The rate of missing information is
  $0.02$ when $\beta=-0.5$, $0.086$ when $\beta=-1.5$, and $0.227$
  when $\beta=-3$.} \\
{\footnotesize \footnotemark[2]: The Monte Carlo error is 0.007.}
\label{toosmallmar}
\end{table}

Next, we consider the scenario where failure indicators may be
missing even if a participant had a negative screener
(i.e. $Q_i=0$). For each such simulation, we randomly selected 40\% of
the observations to have missing failure indicators when
$Q_i=0$. (The mechanism for missing failure indicators when $Q_i=1$
is the same as described previously.) In the first such simulation,
the logistic regression model was correctly specified when
$Q_i=1$. (However, it will be applied to all observations with missing
failure indicators, including those for which $Q_i=0$. Since the
true value of the failure indicator is always 0 when $Q_i=0$, the
model will be biased for these observations.) In the two remaining
simulation scenarios, the model will be misspecified even when $Q_i=1$
by either adding an extra covariate or leaving out a significant
covariate as we did in the earlier simulations.

The results of these three additional simulations are shown in Tables
\ref{mcarweird}, \ref{altmodelmcar}, and \ref{toosmallmcar}. The model
performs well in two of the three scenarios, indicating that our
methodology is robust against misspecification of the logistic
regression model. However, when an important covariate is not included
in the model, the estimates are badly biased. Empirical coverage
ranged from 0\% to 50\%, significantly below the nominal rate. This
indicates that our method can give incorrect results if the predictive
accuracy of the logistic regression model is poor. Note that the
method of Cook and Kosorok \cite{kosorok} also performs poorly in this
scenario. If one cannot accurately estimate which failure indicators
are missing, it is unlikely that any method can produce valid
confidence intervals for the Cox regression coefficients.



\renewcommand{\tablename}{Table}
\begin{table}[h!]
\caption{Results when the Logistic Regression Model
  is Applied to Observations with $Q_i=0$}
\begin{center}
\begin{tabular}{clrccclcl}
\toprule
$\beta$\footnotemark[1]&Method&Bias&SE (Bias)&Width&SE (Width)&Coverage\footnotemark[2]&Running Time (s.)\\ \hline
\midrule
-0.5&Full Data&-0.0029&0.0006&0.1669&0.0001&0.951&0.007\\
&Complete Case&0.0422&0.0007&0.2164&0.0001&0.864&0.006\\
&Treat all as Censored&0.1042&0.0007&0.2132&0.0001&0.516&0.007\\
&Treat all as Failure&0.0883&0.0005&0.1526&0.0001&0.372&0.007\\
&Cook \& Kosorok&-0.002&0.0006&0.1781&0.0002&0.945&21.41\\
&Multiple Imputation&0.0028&0.0006&0.1779&0.0001&0.951&0.48\\ \hline
-1.5&Full Data&-0.0031&0.001&0.3187&0.0002&0.956&0.007\\
&Complete Case&0.053&0.0015&0.4306&0.0004&0.907&0.006\\
&Treat all as Censored&0.12&0.0014&0.4224&0.0003&0.779&0.007\\
&Treat all as Failures&0.8199&0.0007&0.204&0.0001&0&0.007\\
&Cook \& Kosorok&-0.0026&0.0011&0.3483&0.0004&0.948&17.785\\
&Multiple Imputation&0.0128&0.0011&0.3557&0.0006&0.961&0.38\\ \hline
-3&Full Data&-0.0244&0.0026&0.7602&0.0009&0.947&0.008\\
&Complete Case&0.0024&0.0035&1.0686&0.0017&0.954&0.006\\
&Treat all as Censored&0.1043&0.0035&1.0379&0.0015&0.914&0.008\\
&Treat all as Failures&2.5176&0.0008&0.2218&0.0001&0&0.007\\
&Cook \& Kosorok&-0.0104&0.0029&0.9885&0.0028&0.96&18.69\\
&Multiple Imputation&0.0909&0.0029&1.0932&0.0053&0.969&0.424\\
\bottomrule
\hline\\
\end{tabular}
\end{center}
{\footnotesize \footnotemark[1]: The rate of missing information is
  $0.126$ when $\beta=-0.5$, $0.212$ when $\beta=-1.5$, and $0.492$
  when $\beta=-3$.} \\
{\footnotesize \footnotemark[2]: The Monte Carlo error is 0.007.}
\label{mcarweird}
\end{table}

\renewcommand{\tablename}{Table}
\begin{table}
\caption{Results when an Extra Covariate is Included
in the Logistic Regression Model and the Model is Applied to
Observations with $Q_i=0$}
\begin{center}
 \begin{tabular}{clrccclcl}
\toprule
$\beta$\footnotemark[1]&Method&Bias&SE (Bias)&Width&SE (Width)&Coverage\footnotemark[2]&Running Time (s.)\\ \hline
\midrule
-0.5&Full Data&-0.0011&0.0005&0.1669&0.0001&0.956&0.008\\
&Complete Case&0.0442&0.0007&0.2167&0.0001&0.876&0.006\\
&Treat all as Censored&0.1061&0.0007&0.2133&0.0001&0.506&0.007\\
&Treat all as Failures&0.0913&0.0005&0.1525&0.0001&0.342&0.008\\
&Cook \& Kosorok&0.0011&0.0006&0.1783&0.0002&0.959&22.05\\
&Multiple Imputation&0.0064&0.0006&0.1791&0.0002&0.956&0.5\\ \hline
-1.5&Full Data&-0.003&0.001&0.3185&0.0002&0.949&0.007\\
&Complete Case&0.0525&0.0014&0.4288&0.0004&0.911&0.006\\
&Treat all as Censored&0.1171&0.0014&0.4214&0.0003&0.814&0.007\\
&Treat all as Failures&0.8186&0.0007&0.2041&0.0001&0&0.008\\
&Cook \& Kosorok&-0.0022&0.0011&0.3495&0.0004&0.947&18.37\\
&Multiple Imputation&0.0149&0.0011&0.3588&0.0008&0.953&0.397\\ \hline
-3&Full Data&-0.0098&0.0025&0.7555&0.0008&0.943&0.008\\
&Complete Case&0.0042&0.0036&1.0687&0.0017&0.948&0.007\\
&Treat all as Censored&0.116&0.0033&1.0332&0.0014&0.918&0.008\\
&Treat all as Failures&2.5152&0.0008&0.2218&0.0001&0&0.008\\
&Cook \& Kosorok&-0.0013&0.0028&0.9876&0.0028&0.964&20.22\\
&Multiple Imputation&0.1024&0.0028&1.0969&0.0055&0.956&0.46\\
\bottomrule
\hline \\
\end{tabular}
\end{center}
{\footnotesize \footnotemark[1]: The rate of missing information is
  $0.125$ when $\beta=-0.5$, $0.23$ when $\beta=-1.5$, and $0.526$
  when $\beta=-3$.} \\
{\footnotesize \footnotemark[2]: The Monte Carlo error is 0.007.}
\label{altmodelmcar}
\end{table}

\renewcommand{\tablename}{Table}
\begin{table}
\caption{Results when a Relevant Covariate is not
Included in the Logistic Regression Model and the Model is Applied to
Observations where $Q_i=0$}
\begin{center}
\begin{tabular}{clrccclcl}
\toprule
$\beta$\footnotemark[1]&Method&Bias&SE (Bias)&Width&SE (Width)&Coverage\footnotemark[2]&Running Time (s.)\\ \hline
\midrule
-0.5&Full Data&-0.0027&0.0005&0.1667&0.0001&0.963&0.008\\
&Complete Case&0.0418&0.0007&0.2159&0.0001&0.878&0.006\\
&Treat all as Censored&0.1018&0.0007&0.2128&0.0001&0.533&0.008\\
&Treat all as Failures&0.0887&0.0005&0.1523&0.0001&0.375&0.007\\
&Cook \& Kosorok&0.0814&0.0005&0.157&0.0001&0.5&19.66\\
&Multiple Imputation&0.0816&0.0005&0.1573&0.0001&0.471&0.39\\ \hline
-1.5&Full Data&-0.0062&0.001&0.3185&0.0002&0.947&0.007\\
&Complete Case&0.0517&0.0014&0.4294&0.0004&0.903&0.006\\
&Treat all as Censored&0.1154&0.0014&0.4218&0.0003&0.806&0.007\\
&Treat all as Failures&0.8183&0.0007&0.2044&0.0001&0&0.007\\
&Cook \& Kosorok&0.5006&0.0012&0.3808&0.0006&0.004&16.76\\
&Multiple Imputation&0.5151&0.0012&0.3916&0.001&0.005&0.32\\ \hline
-3&Full Data&-0.0126&0.0025&0.756&0.0008&0.952&0.008\\
&Complete Case&0.0057&0.0036&1.0711&0.0016&0.947&0.007\\
&Treat all as Censored&0.1073&0.0035&1.0376&0.0014&0.902&0.008\\
&Treat all as Failures&2.5132&0.0009&0.2218&0.0001&0&0.007\\
&Cook \& Kosorok&0.8862&0.0036&1.0736&0.0028&0.2&17.01\\
&Multiple Imputation&0.9962&0.0036&1.1777&0.0043&0.117&0.353\\
\bottomrule
\hline \\
\end{tabular}
\end{center}
{\footnotesize \footnotemark[1]: The rate of missing information is
  $0.048$ when $\beta=-0.5$, $0.611$ when $\beta=-1.5$, and $0.825$
  when $\beta=-3$.} \\
{\footnotesize \footnotemark[2]: The Monte Carlo error is 0.007.}
\label{toosmallmcar}
\end{table}

\subsection{Simulations Under MNAR} \label{simmnar}
We examined two possible scenarios where the data is MNAR:
\begin{enumerate}
\item In the first, we set 30\% of the censored observations
and 50\% of the failures to have missing indicators.\label{mnar3050}
\item In the second, we set 20\% of the censored observations
and 60\% of the failures to have missing indicators.\label{mnar2060}
\end{enumerate}

The results of these simulations are shown in Tables \ref{mnar3050}
and \ref{mnar2060}. Bias increased for all methods under both MNAR
scenarios.  In particular, the complete case method consistently
displayed a high amount of bias and did not achieve the desired
coverage rate.  For our imputation method and the method of Cook and
Kosorok \cite{kosorok}, the bias and width of the 95\% confidence
interval increased and as the absolute value of the true parameter
value increased. This indicates that when the MAR assumption is
violated, our method as well as the method of Cook and Kosorok
\cite{kosorok} may not be valid. On the other hand, even when the data
was not MAR, our method provided an improvement in terms of bias and
coverage compared to the complete case method and the method that
treats all missing subjects as failures. Moreover, the coverage
probability was slightly greater for our method than for the method of
Cook and Kosorok \cite{kosorok}.

\renewcommand{\tablename}{Table}
\begin{table}
\caption{Simulation Results for MNAR, scenario \ref{mnar3050}}
\begin{center}
\begin{tabular}{clrccclcl}
\toprule
$\beta$\footnotemark[1]&Method&Bias&SE (Bias)&Width&SE (Width)&Coverage\footnotemark[2]&Running Time (s.)\\ \hline
\midrule
-0.5&Full Data&0.0022&0.0006&0.1668&0.0001&0.932&0.007\\
&Complete Case&-0.0695&0.0008&0.2419&0.0001&0.787&0.006\\
&Treat all as Censored&0.0005&0.0008&0.2422&0.0001&0.943&0.007\\
&Treat all as Failures&0.0063&0.0006&0.1702&0.0001&0.929&0.007\\
&Cook \& Kosorok&-0.0014&0.0006&0.1768&0.0001&0.937&19.9\\
&Multiple Imputation&-0.0011&0.0006&0.175&0.0001&0.936&0.46\\ \hline
-1.5&Full Data&-0.0019&0.001&0.3182&0.0002&0.965&0.008\\
&Complete Case&-0.1799&0.0016&0.4704&0.0004&0.699&0.007\\
&Treat all as Censored&-0.0042&0.0016&0.4624&0.0004&0.95&0.007\\
&Treat all as Failures&0.0625&0.001&0.3173&0.0002&0.866&0.007\\
&Cook \& Kosorok&-0.0197&0.0011&0.3505&0.0004&0.934&16.69\\
&Multiple Imputation&-0.0173&0.0011&0.3447&0.0003&0.959&0.38\\ \hline
-3&Full Data&-0.0208&0.0025&0.7571&0.0008&0.946&0.007\\
&Complete Case&-0.3316&0.004&1.1528&0.002&0.846&0.007\\
&Treat all as Censored&-0.0423&0.0038&1.1107&0.0017&0.952&0.007\\
&Treat all as Failures&0.5224&0.0026&0.648&0.0007&0.18&0.008\\
&Cook \& Kosorok&-0.0677&0.0029&0.9291&0.0019&0.922&16.32\\
&Multiple Imputation&-0.0507&0.0028&0.8641&0.0014&0.959&0.4\\
\bottomrule
\hline\\
\end{tabular}
\end{center}
{\footnotesize \footnotemark[1]: The rate of missing information is
  $0.039$ when $\beta=-0.5$, $0.088$ when $\beta=-1.5$, and $0.165$
  when $\beta=-3$.} \\
{\footnotesize \footnotemark[2]: The Monte Carlo error is 0.007.}
\label{mnar3050table}
\end{table}

\renewcommand{\tablename}{Table}
\begin{table}
\caption{Simulation Results for MNAR, scenario \ref{mnar2060}}
\begin{center}
\begin{tabular}{clrccclcl}
\toprule
$\beta$\footnotemark[1]&Method&Bias&SE (Bias)&Width&SE (Width)&Coverage\footnotemark[2]&Running Time (s.)\\ \hline
\midrule
-0.5&Full Data&-0.0022&0.0005&0.167&0.0001&0.952&0.008\\
&Complete Case&-0.0976&0.0009&0.2704&0.0002&0.713&0.007\\
&Treat all as Censored&-0.0017&0.0009&0.2706&0.0001&0.936&0.007\\
&Treat all as Failures&0.0009&0.0006&0.1706&0.0001&0.95&0.008\\
&Cook \& Kosorok&-0.0091&0.0006&0.1796&0.0001&0.944&20.85\\
&Multiple Imputation&-0.0087&0.0006&0.178&0.0001&0.948&0.44\\ \hline
-1.5&Full Data&-0.0007&0.0011&0.3183&0.0002&0.95&0.007\\
&Complete Case&-0.2321&0.0018&0.5311&0.0005&0.586&0.007\\
&Treat all as Censored&-0.0019&0.0017&0.5175&0.0004&0.952&0.007\\
&Treat all as Failures&0.0444&0.0011&0.3202&0.0002&0.9&0.007\\
&Cook \& Kosorok&-0.0397&0.0012&0.3575&0.0004&0.9&16.03\\
&Multiple Imputation&-0.0382&0.0012&0.3537&0.0003&0.932&0.36\\ \hline
-3&Full Data&-0.0112&0.0025&0.7565&0.0009&0.948&0.007\\
&Complete Case&-0.409&0.0043&1.3088&0.0023&0.827&0.007\\
&Treat all as Censored&-0.0255&0.0041&1.2435&0.002&0.954&0.007\\
&Treat all as Failures&0.3801&0.0026&0.682&0.0007&0.425&0.007\\
&Cook \& Kosorok&-0.1038&0.003&0.9663&0.0021&0.919&15.45\\
&Multiple Imputation&-0.0891&0.0028&0.8839&0.0015&0.959&0.38\\
\bottomrule
\hline\\
\end{tabular}
\end{center}
{\footnotesize \footnotemark[1]: The rate of missing information is
  $0.063$ when $\beta=-0.5$, $0.108$ when $\beta=-1.5$, and $0.152$
  when $\beta=-3$.} \\
{\footnotesize \footnotemark[2]: The Monte Carlo error is 0.007.}
\label{mnar060table}
\end{table}

\subsection{Simulations for Poisson Regression} \label{simpois}
We performed simulations to evaluate the performance of our method when the desired time-to-event
analysis is a Poisson regression model rather than a Cox model. Poisson
models are commonly used to estimate incidence rates, which is an
objective of
the OPPERA study.

The simulations were identical to those described in Section
\ref{s:sims} except that the imputed data was used to fit Poisson
regression models rather than
Cox proportional hazards models.  That is, we fit the data from
imputations $j=1,\ldots,m$ to the model
\begin{equation} \label{poissonmodel}
\log(\mu_i)=\alpha+\beta Z_i+\log(V_i).
\end{equation}
where $\mu_i$ is the expected number of cases
and the offset, $\log(V_i)$, is the logarithm of the survival time.
We measured the bias, defined as $\hat{\beta}$ minus the true value, for
$\beta\in\{-0.5,-1.5,-3\}.$

The Cook and Kosorok \cite{kosorok} method does not immediately
generalize to Poisson regression. Consequently, we only compared our
method to the unachievable ideal of no missing data, the complete case
method, and the two ad hoc methods.

The use of Poisson regression allows us to estimate incidence rates.
For each simulation, we estimated the incidence rate based on the
coefficients of the Poisson regression model in (\ref{poissonmodel}).
Specifically, estimated incidence rates for fixed values of
$Z_i$ are given by
\begin{equation}\label{incidence rates}
\exp(\alpha+\beta Z_i)
\end{equation}
The bias, confidence interval width, and coverage probability of each
method are shown in Table \ref{poissonmar}. We also present the
estimated incidence rates for each quartile of the random variable
$Z_i$ (i.e. the quartiles of the $N(2,1)$ distribution) along with
the true (theoretical) rates for each quartile. See Table
\ref{incidencesims}.

Our method had close to 95\% coverage probability
when Poisson regression was used.
None of the other methods had proper coverage for all of the simulations.
Multiple imputation yielded the least bias of all the methods
besides the unachievable ideal of observing all data. It also
produced more narrow confidence intervals than the complete case method
and the method that treats all missing failure indicators as
censored.

The bias evident in parameter estimation was compounded for incidence rates.
The complete case method and the two ad hoc methods consistently
underestimated incidence.
In fact, the complete case method underestimated incidence by about 30-200\%.
By contrast, our method differed from the unachievable ideal by only
about 4-6\%.

\renewcommand{\tablename}{Table}
\begin{table}
\caption{Simulation Results for Poisson Models, MAR}
\begin{center}
\begin{tabular}{clrccclc}
\toprule
$\beta$\footnotemark[1]&Method&Bias&SE (Bias)&Width&SE (Width)&Coverage\footnotemark[2]\\ \hline
\midrule
-0.5&Full Data&-0.0036&0.0005&0.1596&0.0001&0.956\\
&Complete Case&-0.0101&0.0007&0.2067&0.0001&0.948\\
&Treat all as Censored&0.0813&0.0007&0.2049&0.0001&0.652\\
&Treat all as Failures&-0.0002&0.0006&0.1628&0.0001&0.957\\
&Multiple Imputation&-0.0036&0.0006&0.1652&0.0001&0.961\\
\hline
-1.5&Full Data&-0.0005&0.0010&0.2864&0.0002&0.945\\
&Complete Case&-0.1008&0.0015&0.3956&0.0004&0.829\\
&Treat all as Censored&0.0704&0.0014&0.3883&0.0003&0.898\\
&Treat all as Failures&0.0717&0.0010&0.2857&0.0002&0.820\\
&Multiple Imputation&0.0027&0.0011&0.3059&0.0003&0.951\\
\hline
-3&Full Data&-0.0184&0.0021&0.5994&0.0007&0.958\\
&Complete Case&-0.2733&0.0032&0.8710&0.0016&0.792\\
&Treat all as Censored&0.0206&0.0030&0.8485&0.0012&0.954\\
&Treat all as Failures&0.5232&0.0025&0.5544&0.0006&0.085\\
&Multiple Imputation&0.0126&0.0023&0.7012&0.0014&0.964\\
\bottomrule
\hline \\
\end{tabular}
\end{center}
{\footnotesize \footnotemark[1]: The rate of missing information is
  $0.018$ when $\beta=-0.5$, $0.074$ when $\beta=-1.5$, and $0.202$
  when $\beta=-3$.} \\
{\footnotesize \footnotemark[2]: The Monte Carlo error is 0.007.}
\label{poissonmar}
\end{table}

\renewcommand{\tablename}{Table}
\begin{table}[t!]
\caption{Simulation Results for Incidence Rates}
\begin{center}
\begin{tabular}{clrccclc}
\toprule
$\beta$&Method&Q1& SE& Q2 &SE &Q3&SE\\ \hline
\midrule
-0.5&True Rate&0.5154&&0.3679&&0.2626&\\
&Full Data&0.5157&     0.0003 &    0.3671 &    0.0002&     0.2615  &   0.0002\\
&Complete Case&0.4261 &    0.0004  &   0.3019&     0.0002&     0.2142 &    0.0002\\
&Treat all as Censored&0.2991&     0.0003&     0.2254   &  0.0002    & 0.1701&     0.0002\\
&Treat all as Failures&0.4949&     0.0003&     0.3531   &  0.0002   &  0.2521 &    0.0002\\
&Multiple Imputation&0.4902  &   0.0003   &  0.3490     &0.0002    & 0.2486    & 0.0002\\ \hline
-1.5&True Rate&0.1369&&0.0498&&0.0181& \\
&Full Data&0.1375&     0.0001    & 0.0501 &    0.0001 &    0.0183  &       0.0000\\
&Complete Case&0.0968 &    0.0001   &  0.0330  &   0.0001 &    0.0113 &        0.0000\\
&Treat all as Censored&0.0753 &    0.0001 &    0.0288    & 0.0000  &   0.0110&         0.0000\\
&Treat all as Failures&0.1388 &    0.0001&     0.0530    & 0.0001 &    0.0203 &        0.0000\\
&Multiple Imputation&0.1305   &  0.0001&    0.0476     &0.0001 &    0.0174   &      0.0000\\ \hline
-3&True Rate&0.0188&&0.0025&&0.0003&\\
&Full Data&0.0186    & 0.0000   &  0.0025&         0.0000  &   0.0003         &0.0000\\
&Complete Case&0.0106   &  0.0000&     0.0012&         0.0000 &    0.0001         &0.0000\\
&Treat all as Censored&0.0097 &    0.0000    & 0.0013 &        0.0000&     0.0002 &        0.0000\\
&Treat all as Failures&0.0301&     0.0001    & 0.0058&         0.0000 &    0.0011&         0.0000\\
&Multiple Imputation&0.0180 &    0.0000     &0.0025 &        0.0000
& 0.0003&         0.0000\\
\bottomrule
\hline \\
\end{tabular}
\end{center}
\label{incidencesims}
{\footnotesize *: Q1 denotes rates based on the lower quartile of
  $Z_i$.}\\
{\footnotesize *: Q2 denotes rates based on the median of $Z_i$.}\\
{\footnotesize *: Q3 denotes rates based on the upper quartile of
  $Z_i$.}\\
\end{table}

\end{document}